\documentclass[11pt]{article}
\usepackage{a4wide}
\usepackage{amsmath,amssymb}
\usepackage{simplewick}

\usepackage{mathrsfs}
\newcommand{\ms}{\mathscr}

\usepackage{slashed}
\usepackage[dvipdfmx]{graphicx}
\usepackage{color}

\usepackage{comment}
\usepackage[bookmarks=true,bookmarksnumbered=true,setpagesize=false]{hyperref}

\DeclareMathOperator{\diag}{diag}

\newcommand{\ds}{\displaystyle}

\renewcommand{\t}{\text{t}}

\newcommand{\sr}{\stackrel}

\usepackage{enumerate}
\newcommand{\al}[1]{\begin{align}#1\end{align}}
\newcommand{\als}[1]{\begin{align*}#1\end{align*}}
\newcommand{\ov}{\over}
\newcommand{\nn}{\nonumber\\}
\newcommand{\tx}{\text}

\newcommand{\paren}[1]{\left(#1\right)}
\newcommand{\pn}[1]{\left(#1\right)}
\newcommand{\pa}[1]{\left(#1\right)\!{}}
\newcommand{\sqbr}[1]{\left[#1\right]}

\newcommand{\abb}[1]{\left\|#1\right\|}

\newcommand{\fn}[1]{\!\left(#1\right)}

\newcommand{\Pn}[1]{\bigl(#1\bigr)}
\newcommand{\Sqbr}[1]{\bigl[#1\bigr]}

\newcommand{\Fn}[1]{\bigl({#1}\bigr)}

\newcommand{\bs}{\boldsymbol}

\newcommand{\df}{\text{d}}

\newcommand{\I}{\text{I}}

\newcommand{\mc}{\mathcal}

\newcommand{\bmat}[1]{\begin{bmatrix}#1\end{bmatrix}}

\newcommand{\p}{\partial}

\newcommand{\mbf}{\mathbf}
\newcommand{\mbb}{\mathbb}

\newcommand{\h}{\hat}
\newcommand{\ol}{\overline}

\newcommand{\pr}{\prime}

\newcommand{\commutator}[2]{\left[#1\,,\:#2\right]}

\newcommand{\anticommutator}[2]{\left\{#1\,,\:#2\right\}}

\newcommand{\wh}{\widehat}
\newcommand{\wbh}[1]{\boldsymbol{\widehat{#1}}}

\newcommand{\ch}{\check}
\newcommand{\bch}[1]{\boldsymbol{\check#1}}

\newcommand{\sv}{\ell} 

\usepackage{braket}

\usepackage{MnSymbol}
\newcommand{\dK}[1]{\left\lvert#1\right\rrangle}
\newcommand{\dB}[1]{\left\llangle#1\right\rvert}
\newcommand{\dBK}[1]{\left\llangle#1\right\rrangle}
\newcommand{\dBk}[1]{\left\llangle#1\right\rangle}

\newcommand{\anotherautospace}{%
  \mathchoice%
    {\;}
    {\;}
    {\,}
    {\,}
}
\newcommand{\md}{\anotherautospace\middle|\anotherautospace}

\usepackage[normalem]{ulem}

\newcommand{\Nc}{\check N}

\begin{document}

\title{Lorentz-covariant spinor wave packet}

\author{Kin-ya Oda\thanks{E-mail: \tt odakin@lab.twcu.ac.jp} \mbox{}
and Juntaro Wada\thanks{E-mail: \tt wada-juntaro@g.ecc.u-tokyo.ac.jp}}
\maketitle
\begin{center}
\it
$^*$ Department of Mathematics, Tokyo Woman's Christian University, Tokyo 167-8585, Japan\\
$^\dagger$Department of Physics, University of Tokyo, Tokyo 113-0033, Japan\\
\end{center}
\rm
\begin{abstract}\noindent
We propose a novel formulation for a manifestly Lorentz-covariant spinor wave-packet basis. The traditional definition of the spinor wave packet is problematic due to its unavoidable mixing with other wave packets under Lorentz transformations. Our approach resolves this inherent mixing issue. The wave packet we develop constitutes a complete set, enabling the expansion of a free spinor field while maintaining Lorentz covariance. Additionally, we present a Lorentz-invariant expression for zero-point energy.
\end{abstract}

\newpage 
\tableofcontents

\newpage

\section{Introduction}\label{introduction}
In quantum mechanics, wave packets serve as a crucial conceptual and mathematical foundation. In real observations, we never encounter idealized plane-wave states, characterized by zero uncertainty in momentum and infinite uncertainty in position. They do not belong to the Hilbert space because of its non-normalizability.
In quantum field theory (QFT), the plane-wave $S$-matrix is traditionally used, but this approach results in divergences due to the squared energy-momentum delta function. This makes the plane-wave $S$-matrix computation more of a mnemonic than a rigorous derivation for observables; see e.g.\ Ref.~\cite{Weinberg:1995mt}.

Wave packet states have been extensively discussed in various contexts of particle physics phenomenology, including anomalies in vector meson decay~\cite{Ishikawa:2023bnx}, corrections to Fermi's golden rule~\cite{Ishikawa:2019nes,Ushioda:2019hje}, searches for dark photons~\cite{Demidov:2018odn}, applications in quantum computation~\cite{Jordan:2011ci,Jordan:2012xnu,Davoudi:2024wyv}, and studies of neutrino oscillation~\cite{Giunti:1997wq,Cardall:1999ze, Akhmedov:2008jn,Kopp:2009fa,Akhmedov:2009rb,  Akhmedov:2010ua, Akhmedov:2010ms, Wu:2010yr, Akhmedov:2012uu, Blasone:2021cau,Akhmedov:2022bjs,Mitani:2023hpd}.

Despite these applications, theoretical efforts to construct QFT based on wave-packet states have been limited. Previous work has employed Gaussian formalism, utilizing Gaussian wave functions as a complete basis for expanding the free one-particle Hilbert space. This approach has revealed phenomena like time-boundary effects, which are unobservable in plane-wave formalism~\cite{Ishikawa:2021bzf}. However, the Gaussian formalism breaks manifest Lorentz covariance, necessitating a refinement for both aesthetic and pragmatic reasons.

The absence of manifest Lorentz covariance is unsatisfying, as historically, physics has advanced through symmetry-based formalisms, such as the Dirac~\cite{Dirac:1928hu} and Becchi-Rouet-Stora-Tyutin (BRST)~\cite{Becchi:1974xu,Becchi:1975nq,Tyutin:1975qk} formalisms. Therefore, it is desirable to develop a manifestly Lorentz-covariant wave packet formalism.

Practically, the Gaussian formalism complicates calculations due to the lack of Lorentz covariance, evident in the difficulty of deriving explicit analytic formulas for Gaussian wave functions in position space. In contrast, our previous study demonstrated the expansion of scalar fields in QFT using a complete basis of Lorentz-invariant wave packets, facilitating the derivation of analytic formulas for Lorentz-invariant wave functions in scalar fields~\cite{Oda:2021tiv}. Also, the Gaussian wave packet is shown to be a non-relativistic limit of the Lorentz-invariant wave packet for the scalar fields~\cite{Oda:2021tiv}.

In this study, we investigate the wave packet basis for spinors within a Lorentz-covariant framework. Traditional methodologies, including Gaussian formalism (refer to Appendix~A in Ref.~\cite{Ishikawa:2018koj} for an overview) and other approaches towards Lorentz-covariant spinor wave packets~\cite{Naumov:2009zza,Naumov:2010um,Naumov:2013uia,Naumov:2020yyv,Naumov:2022kwz}, assume that the spin dependence of spinor wave packets aligns with that of plane waves. However, this assumption introduces a significant complication: it leads to a complex Lorentz transformation law, which unexpectedly intertwines wave-packet states with different central momenta and positions. Such a phenomenon is physically paradoxical. Imagine a scenario where a single particle is depicted by a wave packet with a distinct central momentum and position. The blending of this wave packet with others having varying central momenta and positions effectively results in an unphysical merging of distinct particles under Lorentz transformation, which is a challenging notion for conventional theories.

In response to this issue, this paper introduces a novel definition of the spinor wave packet that circumvents this problematic mixing: $\dK{X,P,S;\sigma}$, where $X$ and $P$ are its position and momentum centers, $S$ its spin, and $\sigma$ its spatial width-squared. The wave packet $\dK{X,P,S;\sigma}$ is defined by
\al{
\dBK{p,s\md X,P,S;\sigma}
    &\propto
        e^{-ip\cdot\pn{X+i\sigma P}}\ol u\fn{p,s}u\fn{P,S}\nn
    &\quad=   -e^{-ip\cdot\pn{X+i\sigma P}}\ol v\fn{P,S}v\fn{p,s},
      \label{main proposal}
}
where $\dK{p,s}$ is a Lorentz-friendly plane-wave basis state with the momentum $p$ and spin $s$;
further details will be given in subsequent sections, especially in Eq.~\eqref{general choice}.

Our definition ensures that wave packets remain separate and distinct under Lorentz transformations. We then establish the completeness of this newly defined spinor wave packet within the free one-particle subspace. Expanding on this concept, we demonstrate that the spinor field can be effectively expanded using our spinor wave packet. Additionally, we explore the application of this approach to several well-established operators in the wave packet basis, providing a comprehensive understanding of its implications in quantum field theory.

The paper is organized as follows: Section~\ref{Lorentz-covariant spinor wave packet} introduces the new Lorentz-covariant spinor wave-packet basis in the one-particle subspace. Section~\ref{Spinor field expanded by wave packets} extends this to the creation and annihilation operators, showing how the free fermion field can be expanded using this basis. Finally, Section~\ref{Energy, momentum, and charge} presents the expression of several QFT operators in terms of wave packets.

\section{Lorentz-covariant spinor wave packet}
\label{Lorentz-covariant spinor wave packet}
In this section, we point out that the known representation of a spinor wave packet suffers from mixing with other wave packets under Lorentz transformations, and propose a complete set of Lorentz-covariant spinor wave-packet basis without the difficulty of mixing.

We work in the $\pn{d+1}$-dimensional Minkowski space $\mbf M^{d+1}$ spanned by coordinate system $x=\pn{x^0,\bs x}=\pn{x^0,x^1,\dots,x^d}\in\mbb R^{1,d}$, with $d=3$ spatial dimensions.
We take the almost-plus metric signature $\pn{-,+,\dots,+}$; expressions in the opposite convention can be found in Appendix~\ref{metric conversion}.
We only consider a massive field, $m>0$, and always take $\pn{d+1}$-momenta on-shell,
$p^0=\sqrt{m^2+\bs p^2}$,
throughout this paper unless otherwise stated.
When an on-shell momentum appears in an argument of a function such as $f\fn{\bs p}$, we use both $d$ and $\pn{d+1}$-dimensional notations interchangeably: $f\fn{\bs p}=f\fn{p}$.

\subsection{Spinor plane waves, revisited}
To spell out our notation, we summarize basic known facts on the spinor plane waves.
A free Dirac field $\wh\psi\fn{x}$ can be expanded by plane wave as follows,
\al{
\wh\psi\fn{x}
	&=	\sum_s \int{\df^d\bs p\ov2p^0}\pn{
			u(p,s){e^{ip\cdot x}\ov\pn{2\pi}^{d/2}}\wh\alpha\fn{p,s}
			+v(p,s){e^{-ip\cdot x}\ov\pn{2\pi}^{d/2}}\wh \beta^\dagger\fn{p,s}},
	\label{Dirac field expansion plane wave}
}
where $u\fn{p,s}$ and $v\fn{p,s}$ are plane-wave solutions of the Dirac equation
\al{
&(i\slashed{p}+m)u(p,s)=0,\nn
&(i\slashed{p}-m)v(p,s)=0,
	\label{Dirac equation}
}
with $s=\pm1/2$ being the spin in the rest frame of each solution. 
Throughout this paper, we suppress the spinor indices $a=1,\dots,2^{\left\lfloor\pn{d+1}/2\right\rfloor}$ for $\wh\psi$, $u$, $v$, etc.\ when unnecessary.

These solutions satisfy the following completeness relations,
\al{
\sum_s u(p,s)\ol{u}(p,s)
	&=	-i\slashed{p}+m,\nn
\sum_s v(p,s)\ol{v}(p,s)
	&=	-i\slashed{p}-m,
	\label{completeness of spinor}
}
and their normalization is
\al{
&\ol{u}(p,s)u\fn{p,s'}=2m\delta_{ss'},\nn
&\ol{v}(p,s)v\fn{p,s'}=-2m\delta_{ss'},
    \label{normalization of u and v}
}
where $\ol\psi:=\psi^\dagger\beta$ is the Dirac adjoint.\footnote{
\label{metric convention footnote}
We adopt the spinor notation in Ref.~\cite{Weinberg:1995mt}: $\anticommutator{\gamma^\mu}{\gamma^\nu}=2\eta^{\mu\nu}I$, where $\eta:=\diag\fn{-1,1,\dots,1}$ and $I$ is the unit matrix in the spinor space. Here, $\beta:=i\gamma^0$ is distinguished from the operator $\wh\beta$.
}
The coefficients $\wh\alpha\fn{p,s}$ and $\wh\beta^\dagger\fn{p,s}$ in Eq.~\eqref{Dirac field expansion plane wave} are the annihilation and creation operators
for particle and anti-particle, respectively, that satisfy the following anticommutation relations:
\al{
\anticommutator{\wh\alpha\fn{p,s}}{\wh\alpha^{\dagger}\fn{p',s'}}
	&=	\delta_{ss'}2p^0\delta^d\fn{\bs{p}-\bs{p}'}
	\wh1,\nn
\anticommutator{\wh\beta\fn{p,s}}{\wh\beta^{\dagger}\fn{p',s'}}
	&=	\delta_{ss'}2p^0\delta^d\fn{\bs{p}-\bs{p}'}
	\wh1,\nn
\tx{others}
	&=	0.
	\label{anti-commutator}
}

Free one-particle subspaces of particle and anti-particle are spanned by the following plane-wave bases:
\al{
\wh \alpha^{\dagger}\fn{p,s}\ket{0}
	&=:	\dK{p,s,n},&
\wh \beta^{\dagger}\fn{p,s}\ket{0}
	&=:	\dK{p,s,n^\tx{c}},
 \label{spinor plane waves}
}
where $n$ and $n^\tx{c}$ denote the particle and anti-particle of the species $n$, respectively. The anticommutator~\eqref{anti-commutator} leads to the inner product:
\al{
\dBK{p,s,N\md p',s',N'}
	&:=	2p^0\delta^d\fn{\bs p-\bs p'}\delta_{ss'}\delta_{NN'},
		\label{bases inner product}
}
where $N=n,n^\tx{c}$ labels the particle and anti-particle.
The normalization~\eqref{bases inner product} leads to the completeness relation (resolution of identity) in the free one-particle subspace of each $N=n,n^\tx{c}$:
\al{
\sum_s\int{\df^d\bs p\ov2p^0}\dK{p,s,N}\dB{p,s,N}
	&=	\h1.
	\label{plane-wave with spin completeness}
}

The Lorentz transformation law of the plane wave reads
\al{
\wh U\fn{\Lambda}\dK{p,s}
	=	\sum_{s'}\dK{\Lambda p,s'}D_{s's}\Fn{W\fn{\Lambda,p}},
		\label{Lorentz transformation on plane-wave momentum state}
}
where $D$ is the spin-$s$ representation of the Winger rotation $SO(d)$; see Appendix~\ref{Wigner section} for details.

\subsection{Lorentz-covariant spinor wave packet}
In this subsection, we first briefly review basic facts on Lorentz-invariant scalar wave packets~\cite{Kaiser:1977ys,Kaiser:1978jg}, which is discussed in our previous work~\cite{Oda:2021tiv}.  Next, we point out the difficulty in the conventional treatment of the spinor wave packet~\cite{Naumov:2009zza,Naumov:2010um,Naumov:2013uia,Naumov:2020yyv}. Then, we propose a new definition of the spinor wave packet and show that we can avoid this difficulty in our expression. 

\subsubsection{Brief review of Lorentz-invariant scalar wave packet}\label{LIWP section}
For central position $X$ and momentum $P$ in $\pn{d+1}$-dimensions,
a Lorentz-invariant scalar wave packet $\dK{\Pi}$ is defined by~\cite{Kaiser:1977ys,Kaiser:1978jg}:\footnote{See e.g.\ Refs.~\cite{Naumov:2013uia,Oda:2021tiv} for reviews.}
\al{
\dBK{p\md\Pi}
	&:=	N_\phi e^{-ip\cdot\pn{X+i\sigma P}},
	\label{LIWPdefinition}
}
where $\Pi$ denotes the phase space\footnote{
Here, $\Pi$ includes the wave-packet central time $X^0$.
Though $P^0=\sqrt{m^2+\bs P^2}$ is not an independent variable, we also include it for the convenience of writing its Lorentz transformation below.
}
\al{
\Pi &:=  \pn{X,P},
}
and the normalization factor
\al{
N_\phi
	&:=	{\pn{\sigma\ov\pi}^{d-1\ov4}\ov\sqrt{K_{d-1\ov2}\fn{2\sigma m^2}}}
		\label{normalization factor N}
}
provides $\dBK{\Pi|\Pi}=1$, in which $K_n\fn{z}$ is the modified Bessel function of the second kind. Here and hereafter, we fix $\sigma$ unless otherwise stated.  

The wave function and the inner product are obtained as~\cite{Kaiser:1977ys,Oda:2021tiv}
\al{
\dBK{x\md\Pi}
	&=	N_\phi{m^{d-1}\ov\sqrt{2\pi}}{K_{d-1\ov2}\fn{\abb\xi}\ov\abb\xi^{d-1\ov2}}, 
		\label{scalar wave-packet function}\\
\dBK{\Pi\md\Pi'}
	&=	N_\phi^2\pn{2\pi m^2}^{d-1\ov2}
		{K_{d-1\ov2}\pn{\abb\Xi}\ov\abb\Xi^{d-1\ov2}},
		\label{inner product of wave packet}
}
where for any complex vector $V^\mu$, we write $\abb V:=\sqrt{-V^2}$, namely,
\al{
\abb\xi
	&=	m\sqrt{\sigma^2m^2+\pn{x-X}^2-2i\sigma P\cdot\pn{x-X}},
		\label{xi defined}\\
\abb\Xi
	&=	m\sqrt{\Pn{\pn{X-X'}-i\sigma\pn{P+P'}}^2},
	\label{Xi parameter}
}
with $\xi^\mu:=m\sqbr{\sigma P^\mu+i\pn{x-X}^\mu}$
and $\Xi^\mu:=m\sqbr{\sigma\pn{P+P'}^\mu+i\pn{X-X'}^\mu}$.\footnote{
The abuse of notation is understood such that a vector-squared $V^2:=-\pn{V^0}^2+\bs V^2$ is distinguished from the second component of $V$ by the context.
}
We note that there is no branch-cut ambiguity for the square root as long as $m>0$
~\cite{Oda:2021tiv}.

With this state, the momentum expectation value and its (co)variance become~\cite{Kaiser:1977ys}
\al{
\dBK{\h p^\mu}_{\phi}
	&:=	\int{\df^d\bs p\ov2p^0}\dBK{\Pi\md p}p^\mu\dBK{p\md\Pi}
	=	 \mc M_\phi P^\mu,
			\label{momentum expectation}\\
\dBK{\h p^\mu\h p^\nu}_{\phi}
	&=	{K_{d+3\ov2}\fn{2\sigma m^2}\ov K_{d-1\ov2}\fn{2\sigma m^2}}
                P^\mu P^\nu
        +{\mc M_\phi\ov2\sigma}\eta^{\mu\nu},
		\label{momentum covariance}
}
where 
\al{
\mc M_\phi
    &:= {K_{d+1\ov2}\fn{2\sigma m^2}\ov K_{d-1\ov2}\fn{2\sigma m^2}}.
		\label{M_phi given}
}
In general, a matrix element of $\h p$ becomes
\al{
\dBK{\h p^\mu}_{\Pi,\Pi'}
	&:=	\int{\df^d\bs p\ov2p^0}\dBK{\Pi\md p}p^\mu\dBK{p\md\Pi'} 
	=	\pn{2\pi m^2}^{d-1\ov2}N_\phi^2\,m\Xi^\mu{K_{d+1\ov2}\fn{\abb\Xi}\ov\abb\Xi^{d+1\ov2}}.
		\label{matrix element of p}
}

Let us consider a spacelike hyperplane $\Sigma_{N,T}=\Set{X|N\cdot X+T=0}$ in the space of central position $X$; see Appendix~\ref{``Slanted'' foliation}.
One can write the completeness relation in the position-momentum phase space in a manifestly Lorentz-\emph{invariant} fashion~\cite{Oda:2021tiv} (see also Ref.~\cite{Kaiser:1977ys}):
\al{
\int\df^{2d}\Pi_{\phi}
	\dK{\Pi}\dB{\Pi}
	&=	\h 1,
	\label{completeness relation for scalar}
}
where $\h1$ denotes the identity operator in the one-particle subspace and
the Lorentz-\emph{invariant} phase-space volume element is given by
\al{
\int\df^{2d}\Pi_{\phi}
	&:=	{1\ov\mc M_\phi}
        \int{\df^d\Sigma^\mu_X\ov\pn{2\pi}^d}\pn{-2P_\mu}{\df^d\bs P\ov2P^0},
		\label{Lorentz invariant measure}
}
in which
\al{
\df^d\Sigma^\mu_X
	&:=	\df^{d+1}X\,\delta\fn{N\cdot X+T}N^\mu
    \label{volume element vector}
}
is the Lorentz-covariant volume element.
We stress that $\sigma$ is not summed nor integrated in the identity~\eqref{completeness relation for scalar} and that the identity holds for any fixed $\sigma$.

Let us consider a ``time-slice frame'' $\ch X$ of the central-position space in which $\ch \Sigma_{\ch N,T}$ becomes an equal-time hyperplane $\ch X^0=T$,
\al{
\ch X
    &:= \ms L^{-1}\fn{N}X,
}
where the ``standard'' Lorentz transformation $\ms L\fn{N}$ is defined by $N=:\ms L\fn{N}\ell$, with $\ell$ denoting $\ell:=\pn{1,\bs 0}$ in any frame; note that $\ch N=\ms L^{-1}\fn{N}N=\ell$ by definition; see Appendix~\ref{``Slanted'' foliation} for details.
On the constant-$\ch X^0$ hyperplane $\ch \Sigma_{\Nc,T}=\Set{\ch X|\ch X^0=T}$, the Lorentz-invariant phase-space volume element reduces to the familiar form:
\al{
\int\df^{2d}\ch\Pi_{\phi}
	&=  {1\ov\mc M_\phi}
        \int_{\ch X^0=T}{\df^d \bch X\,\df^d\bch P\ov\pn{2\pi}^d}.
}
Note that $\mc M_\phi\to1$ in the non-relativistic limit $\sigma m^2\gg1$.

\subsubsection{Difficulty in spin-diagonal representation}
In the literature~\cite{Naumov:2009zza,Naumov:2010um,Naumov:2013uia,Naumov:2020yyv} a so to say \emph{spin-diagonal} one-particle wave-packet state $\dK{\Pi,S}_\tx{D}$ with a spin~$S$ has been defined as
\al{
\dBK{p,s\md\Pi,S}_\tx{D}
	&:=	\dBK{p\md\Pi}\delta_{sS}
	\label{trivial choice}
}
where $\dBK{p\md\Pi}$ is nothing but the scalar Lorentz-invariant wave packet~\eqref{LIWPdefinition}.\footnote{
In the literature, the normalization and $X$-dependence~\cite{Oda:2021tiv} have been omitted.
}
Its normalization becomes
\al{
\dBK{\Pi,S\md\Pi',S'}_\tx{D}
	&= 	\sum_s\int{\df^d\bs p\ov2p^0}\dBK{\Pi,S\md p,s}_\tx{D}\dBK{p,s\md\Pi',S'}_\tx{D}\nn
	&=	\sum_s\int{\df^d\bs p\ov2p^0}\dBK{\Pi\md p}\dBK{p\md\Pi'}\nn
	&=	\dBK{\Pi\md\Pi'}\delta_{SS'},
		\label{normalization N of wave packet}
}
where $\dBK{\Pi\md\Pi'}$ is given in Eq.~\eqref{inner product of wave packet}.\footnote{
An inner product of the spin-diagonal wave-packet state and another state $\Ket{\psi}$ is understood as
$\dBk{\Pi,S\md\psi}_\tx{D}:=\pn{\dK{\Pi,S}_\tx{D}}^\dagger\Ket{\psi}=:\dB{\Pi,S}_\tx{D}\Ket{\psi}$.
We will never consider an inner product of the spin-diagonal wave-packet state and a phase-space-diagonal wave-packet state that appears below so that this notation will not cause confusion.
}
This leads to the following completeness relation in the one-particle subspace
\al{
\sum_S\int\df^{2d}\Pi_{\phi}\dK{\Pi,S}_\tx{D}\dB{\Pi,S}_\tx{D}
	&=	\h1,
}
generalizing the completeness relation of the scalar wave packet~\eqref{completeness relation for scalar}.

Once the wave-packet state is defined, its Lorentz transformation law is obtained as
\al{
\wh U\fn{\Lambda}\dK{\Pi,S}_\tx{D}
	&=	\sum_s\int{\df^d\bs p\ov2p^0}\wh U\fn{\Lambda}\dK{p,s}\dBK{p,s\md\Pi,S}_\tx{D}\nn
	&= 	\sum_{s,s'}\int{\df^d\bs p\ov2p^0}
			\sum_{S'}\int\df^{2d}\Pi'_{\phi}\dK{\Pi',S'}_\tx{D}\dBK{\Pi',S'\md\Lambda p,s'}_\tx{D}
			D_{s's}\Fn{W\fn{\Lambda,p}}
			\dBK{p,s\md\Pi,S}_\tx{D}\nn
	&= \sum_{S'}\int\df^{2d}\Pi'_{\phi}\dK{\Lambda\Pi',S'}_\tx{D}
		\dBK{D_{S'S}\Fn{W\fn{\Lambda,\h p}}}_{\Pi',\Pi},
		\label{spinor wave packet transformation}
}
where
\al{
\Lambda\Pi:=\pn{\Lambda X,\Lambda P}.
}
and
\al{
\dBK{D_{S'S}\Fn{W\fn{\Lambda,\h p}}}_{\Pi',\Pi} 
	:= \dB{\Pi'}\left(\int{\df^d\bs p\ov2p^0}D_{S'S}\Fn{W\fn{\Lambda,p}}\dK{p}\dB{p} \right)\dK{\Pi}.
}
We see that the spin-diagonal choice~\eqref{trivial choice} leads to the complicated transformation law~\eqref{spinor wave packet transformation} mixing the wave-packet state with the others having various centers of momentum and position.

Below, we will show that we can indeed realize a physically reasonable transformation law, so to say the \emph{phase-space-diagonal} representation, which evades the mixing with other states~\eqref{spinor wave packet transformation}:
\al{
\wh U\fn{\Lambda}\dK{\Pi,S}
		&=\sum_{S'}\dK{\Lambda \Pi,S'}C_{S'S}\fn{\Lambda, \Pi},
		\label{irreducible representation for WPS}
}
where $C_{S'S}\fn{\Lambda, \Pi}$ is a yet unspecified representation function.

\subsubsection{Phase-space-diagonal representation}
Instead of the conventional choice~\eqref{trivial choice}, we propose to define
\al{
\dBK{p,s,N\md\Pi,S,N'}
    &:=   N_\psi e^{-ip\cdot\pn{X+i\sigma P}}M_{sS}\fn{p,P} \delta_{NN'}\nn
    &
    \pn{={N_\psi\ov N_\phi}\dBK{p\md\Pi}M_{sS}\fn{p,P}\delta_{NN'}},
		\label{general choice}
}
where the key element is
\al{
M_{sS}\fn{p,P}
	&:=	
		\ds{\ol u\fn{p,s}u\fn{P,S}\ov2m}
	=	-\ds{\ol v\fn{P,S}v\fn{p,s}\ov2m},
  \label{M defined}
}
and $N_\psi$ is a normalization factor to be fixed below;
in the second step in Eq.~\eqref{M defined}, we used
$u(p,s)=C v^*(p,s)$, with the charge conjugation matrix $C=-\gamma^2$ ($=C^*=C^\t=C^\dagger$) in our notation.
Note that $M_{sS}\fn{p,p}=\delta_{sS}$.
Here and hereafter, for notational simplicity, we omit the label $N$ that distinguishes the particle and antiparticle unless otherwise stated.

The definition~\eqref{M defined} leads to\footnote{
One can show it as
\als{
M_{sS}\fn{p,P}
	&=	{1\ov2m}\ol u\fn{p,s}S^{-1}\fn{\Lambda}S\fn{\Lambda}u\fn{P,S}
	=	{1\ov2m}\sum_{s',S'}D^*_{s's}\Fn{W\fn{\Lambda,p}}\ol u\fn{\Lambda p,s'}u\fn{\Lambda P,S'}D_{S'S}\Fn{W\fn{\Lambda,P}}\nn
	&=	\sum_{s',S'}D^*_{s's}\Fn{W\fn{\Lambda,p}}
		M_{s'S'}\fn{\Lambda p,\Lambda P}D_{S'S}\Fn{W\fn{\Lambda,P}}.
}
}
\al{
M_{sS}\fn{p,P}
    &=  \sum_{s',S'}D^*_{s's}\Fn{W\fn{\Lambda,p}}
		M_{s'S'}\fn{\Lambda p,\Lambda P}D_{S'S}\Fn{W\fn{\Lambda,P}}.
}
Then it follows that
\al{
\dBK{p,s\md\Pi,S}
	&=	\sum_{s',S'}D^*_{s's}\Fn{W\fn{\Lambda,p}}
		\dBK{\Lambda p,s\md\Lambda \Pi,S}D_{S'S}\Fn{W\fn{\Lambda,P}}.
	\label{Lorentz transformation for wave packet wave function}
}
The identity~\eqref{Lorentz transformation for wave packet wave function} results in\footnote{
This can be shown as
\als{
\wh U\fn{\Lambda}\dK{\Pi,S}
	&=	\sum_s\int{\df^d\bs p\ov2p^0}\wh U\fn{\Lambda}\dK{p,s}\dBK{p,s\md\Pi,S}
	=	\sum_{s',S'}\int{\df^d\bs p\ov2p^0}
			\dK{\Lambda p,s'}\dBK{\Lambda p,s'\md\Lambda\Pi,S'}D_{S'S}\Fn{W\fn{\Lambda,P}}\nn
	&=	\sum_{S'}\dK{\Lambda\Pi,S'}D_{S'S}\Fn{W\fn{\Lambda,P}},
}
where we have used Eqs.~\eqref{Lorentz transformation on plane-wave momentum state} and \eqref{Lorentz transformation for wave packet wave function} and then the unitarity~\eqref{D unitarity} in the second equality.
}
\al{
\wh U\fn{\Lambda}\dK{\Pi,S}
    &=  \sum_{S'}\dK{\Lambda\Pi,S'}D_{S'S}\Fn{W\fn{\Lambda,P}}.
}
As promised, we have realized the phase-space-diagonal representation~\eqref {irreducible representation for WPS}.

Now we show that the normalization $\dBK{\Pi,S\md\Pi,S}=1$ is realized by the choice
\al{
\label{normalization constant for spinor}
N_\psi=
      \sqrt{{2\ov1+\mc M_\phi}} N_{\phi},
}
where $N_\phi$ is given in Eq.~\eqref{normalization factor N}.
Let us first compute
\al{
\dBK{\Pi,S\md\Pi,S}
	&=	\sum_{s}\int{\df^d\bs p\ov2p^0}\dBK{\Pi,S\md p,s}\dBK{p,s\md\Pi,S}\nn
	&=	{N_\psi^2\ov\pn{2m}^2}\ol u\fn{P,S}\pn{\int{\df^d\bs p\ov2p^0}{\dBK{\Pi\md p}\dBK{p\md\Pi}\ov N_\phi^2}
		\pn{-i\slashed p+m}}
		u\fn{P,S}\nn
	&=	{N_\psi^2\ov\pn{2m}^2}\ol u\fn{P,S}{\dBK{-i\slashed{\h p}+m}_\phi\ov N_\phi^2}
		u\fn{P,S}\nn
    &\pn{=  {N_\psi^2\ov\pn{2m}^2}\ol v\fn{P,S}{\dBK{-i\slashed{\h p}-m}_\phi\ov N_\phi^2}
		v\fn{P,S}},
  \label{normalization for particle}
}
where we used Eq.~\eqref{completeness of spinor} in the second line;
we used $\pn{\beta C}^2=-1$ and $\beta C\gamma^{\t\mu}\beta C=\gamma^\mu$ in the last line, which is convenient for the antiparticle.
The expectation value $\dBK{\h p^{\mu}}_\phi$ is presented in Eq~\eqref{momentum expectation}, from which we get
\al{
\label{pslashexpectation}
\dBK{-i\slashed{\h p}\pm m}_{\phi}
    =   -i\mc M_\phi\slashed P
        \pm m.
}
Therefore, using the Dirac equation~\eqref{Dirac equation} and then the normalization~\eqref{normalization of u and v}, we see that the choice~\eqref{normalization constant for spinor} provides the normalized state.

Finally, the inner product is given by the same procedure:
\al{
\dBK{\Pi,S,N\md\Pi',S',N'}
	&=	
        \delta_{NN'}{N_\psi^2\ov\pn{2m}^2}\ol u\fn{P,S}{\dBK{-i\slashed{\h p}+m}_{\Pi,\Pi'}\ov N_\phi^2}
		u\fn{P',S'}\nn
  &\pn{=
        \delta_{NN'}{N_\psi^2\ov\pn{2m}^2}\ol v\fn{P,S}{\dBK{-i\slashed{\h p}-m}_{\Pi,\Pi'}\ov N_\phi^2}
		v\fn{P',S'}
  },
 \label{inner product of Lorentz covariant wave pacekt}
}
where, $\dBK{-i\slashed{\h p}\pm m}_{\Pi,\Pi'}=-i\dBK{\slashed{\h p}}_{\Pi,\Pi'}\pm m\dBK{\Pi\md\Pi'}$; see Eqs.~\eqref{inner product of wave packet} and \eqref{matrix element of p}.
Hereafter, we adopt this representation for the Lorentz-covariant spinor wave packet.

\subsection{Momentum expectation value}
In this subsection, we compute the momentum expectation value of the Lorentz covariant spinor wave packet:
\al{
\dBK{\h p^{\mu}}_{\psi} 
	&:= \sum_s \int{\df^d\bs p\ov2p^0}\dBK{\Pi,S\md p,s}p^\mu\dBK{p,s\md \Pi,S}.
}
This will be an important parameter in the following.
Putting Eq.~\eqref{general choice}, we obtain
\al{
\dBK{\h p^{\mu}}_{\psi} 
	&={N_\psi^2\ov 4m^2}\sum_s \int{\df^d\bs p\ov2p^0}{\dBK{\Pi\md p}p^\mu\dBK{p\md\Pi}\ov N_\phi^2}\ol{u}(P,S)u(p,s)\ol{u}(p,s)u(P,S)\nn
	&= {N_\psi\ov 4m^2}\ol{u}(P,S){\dBK{\h p^{\mu}(-i\slashed{\h p}+m)}_{\phi}\ov N_\phi^2}u(P,S)\ 
 \pn{
    ={N_\psi\ov 4m^2}\ol{v}(P,S){\dBK{\h p^{\mu}(-i\slashed{\h p}-m)}_{\phi}\ov N_\phi^2}v(P,S)
        },
        \label{p expectation value by psi}
}
where we used Eq.~\eqref{completeness of spinor}. The expectation value and its covariance $\dBK{\h p^{\mu}}_{\phi}$, $\dBK{\h p^{\mu} \h p^{\nu}}_{\phi}$ are shown in Eqs~\eqref{momentum expectation} and \eqref{momentum covariance}. Thus,
\al{
\label{pslashexpectation2}
\dBK{\h p^{\mu}(-i\slashed{\h p}\pm m)}_{\phi}
    &=  -i \paren{
			{K_{d+3\ov2}\fn{2\sigma m^2}\ov K_{d-1\ov2}\fn{2\sigma m^2}}
                P^{\mu} \slashed{P}
            +{\mc M_\phi\ov2\sigma}\gamma^{\mu}
			} \pm\mc M_\phi m P^{\mu}.
}
Hence, using the Dirac equation~\eqref{Dirac equation} and then the normalization~\eqref{normalization of u and v}, we get
\al{
{1\ov4m^2}\ol{u}(P,S)\dBK{\h p^{\mu}(-i\slashed{\h p}+m)}_{\phi}u(P,S)
    &=  {1\ov2}\paren{
			\mc M_\phi +
			{\mc M_\phi\ov2\sigma m^2}
			+
				{K_{d+3\ov2}\fn{2\sigma m^2}\ov K_{d-1\ov2}\fn{2\sigma m^2}}}P^{\mu},\nn
{1\ov4m^2}\ol{v}(P,S)\dBK{\h p^{\mu}(-i\slashed{\h p}-m)}_{\phi}v(P,S)
    &=  {1\ov2}\paren{
			\mc M_\phi +
			{\mc M_\phi\ov2\sigma m^2}
			+
				{K_{d+3\ov2}\fn{2\sigma m^2}\ov K_{d-1\ov2}\fn{2\sigma m^2}}}P^{\mu},
\label{useful equation}
}
Therefore, the momentum expectation value is given by
\al{
\label{momentum expectation for spinor}
\dBK{\h p^{\mu}}_{\psi} 
	&=\mc M_\psi P^{\mu},
}
where
\al{
\label{M_psi given}
\mc M_\psi
    &:=  {1\ov1+\mc M_\phi}\sqbr{
            {K_{d+3\ov2}\fn{2\sigma m^2}\ov K_{d-1\ov2}\fn{2\sigma m^2}}
            +\mc M_\phi\pn{1 +{1\ov2\sigma m^2}}
            }.
}
Note that $\mc M_\psi\to1$ in the non-relativistic limit $\sigma m^2\gg1$.

\subsection{Completeness}
In this subsection, we will prove the following completeness relation for Lorentz-covariant spinor wave packet,
\al{
\sum_S\int\df^{2d}\Pi_{\psi}\dK{\Pi,S}\dB{\Pi,S}
	&=	\h1,
		\label{completeness of spin wave packet}
}
where
\al{
\label{Lorentz invariant measure for spinor}
\int\df^{2d}\Pi_{\psi}
	&:=
        {1\ov\mc M_\psi}
        \int_{\Sigma_{N,T}}{\df^d\Sigma^\mu_X\ov\pn{2\pi}^d}\pn{-2P_\mu}{\df^d\bs P\ov2P^0}\nn	
	&=  {\mc M_\phi\ov\mc M_\psi}
        \int\df^{2d}\Pi_{\phi},
}
in which $\df^{2d}\Pi_{\phi}$, $\mc M_\phi$, and $\mc M_\psi$ are given in Eqs.~\eqref{Lorentz invariant measure},~\eqref{M_phi given}, and~\eqref{M_psi given} respectivity.

To prove Eq.~\eqref{completeness of spin wave packet}, we rewrite it as a matrix element for both-hand sides, sandwiched by the plane-wave bases~\eqref{spinor plane waves}:
\al{
&\quad{N_\psi^2\ov\mc M_\psi}\int_{\Sigma_{N,T}}{\df^{d+1}X\ov\pn{2\pi}^d}\delta\fn{N\cdot X+T}(-2P\cdot N){\df^d\bs P\ov2P^0}{\dBK{p\md\Pi}\dBK{\Pi\md q}\ov N_\phi^2}&\nn
&\qquad\times \ds{1\ov4m^2}\sum_S \ol u\fn{p,s}u\fn{P,S} \ol u\fn{P,S}u\fn{q,s'}\nn
	&=	2 p^0 \delta^d (\bs{p}-\bs{q})\delta_{ss'},
	\label{expression of completeness of spin wave packet}
}
where we used Eq.~\eqref{bases inner product} on the right-hand side.
On the left-hand side, we integrate $X$ over $\Sigma_{N,T}$ by exploiting its Lorentz invariance, choosing a coordinate system where it becomes a constant-$\ch X^0$ hyperplane $\ch \Sigma_{\Nc,T}$ with $\ch X^0=T$. Then left-hand side in Eq.~\eqref{expression of completeness of spin wave packet} becomes
\al{
(\tx{l.h.s.})
    &= {N_{\psi}^2\ov\mc M_\psi^2}\delta^d (\bs{p}-\bs{q})\int{\df^d\bs P\ov2P^0}2P^0{e^{2\sigma P \cdot p}\ov N_\phi^2}
        {\ol u\fn{p,s} \pn{-i\slashed P+m} u\fn{p,s'}\ov4m^2} \nn
	&= {N_{\psi}^2\ov\mc M_\psi^2}  \delta^d (\bs{p}-\bs{q}) \ol u\fn{p,s}{\dBK{2\h p^0(-i\slashed{\h p}+m)}_{\phi}\ov N_\phi^2}
		u\fn{q,s'}\nn
	&=2 p^0 \delta^d (\bs{p}-\bs{q}),
 \label{lhs eq}
} 
where we used Eq.~\eqref{completeness of spinor} in the first line, and  Eqs.~\eqref{useful equation} and~\eqref{M_psi given} in the last line. Thus, Eq.~\eqref{expression of completeness of spin wave packet}, and hence the completeness~\eqref{completeness of spin wave packet}, is proven.

\section{Spinor field expanded by wave packets}
\label{Spinor field expanded by wave packets}
Now we define the creation and annihilation operators of the Lorentz-covariant wave packet.
We write a free spin-$1/2$ one-particle state of $n$th spinor particle $\dK{\Pi,S;n}$ and of its anti-particle $\dK{\Pi,S;n^{\tx c}}$.
Similarly to the plane wave case, we define wave-packet creation operators by
\al{
\wh A^{\dagger}\fn{\Pi,S}\ket{0}
	&:=	\dK{\Pi,S,n},\\
\wh B^{\dagger}\fn{\Pi,S}\ket{0}
	&:=	\dK{\Pi,S,n^\tx{c}},
}
and annihilation operators $\wh A\fn{\Pi,S},\wh B\fn{\Pi,S}$ by their Hermitian conjugate, with mass dimensions $\sqbr{\wh A^{\dagger}\fn{\Pi,S}}=\Sqbr{\dK{\Pi,S;n}}=0$, etc.
Then, the completeness relation~\eqref{completeness of spin wave packet} on the one-particle subspace reads
\al{
\Bra{0}\wh\alpha\fn{p,s;n}
	&=	\sum_S\int\df^{2d}\Pi_{\psi}\dBK{p,s\md\Pi,S}\Bra{0}\wh A\fn{\Pi,S},
}
and similarly for the anti-particles.
Then, we can naturally generalize it to an operator relation that is valid on the whole Fock space:
\al{
\wh\alpha\fn{p,s}
	&=	\sum_S \int\df^{2d}\Pi_{\psi}\dBK{p,s\md\Pi,S}\wh A\fn{\Pi,S},
		\label{alpha in terms of A for spinor}\\
\wh\beta\fn{p,s}
	&=	\sum_S \int\df^{2d}\Pi_{\psi}\dBK{p,s\md\Pi,S}\wh B\fn{\Pi,S}.
		\label{beta in terms of B for spinor}
}
Similarly, the completeness of the plane wave~\eqref{plane-wave with spin completeness} leads to the expansion of these creation and annihilation operators:
\al{
\wh A\fn{\Pi,S}
	&=	\sum_s\int{\df^d\bs p\ov2p^0}\dBK{\Pi,S\md p,s}\wh\alpha\fn{p,s},\\
\wh B\fn{\Pi,S}
	&=	\sum_s\int{\df^d\bs p\ov2p^0}\dBK{\Pi,S\md p,s}\wh\beta\fn{p,s}.
}
From the above equations, we can derive the anti-commutation relation of the creation and annihilation operators:
\al{
\anticommutator{\wh A\fn{\Pi,S}}{\wh A^\dagger\fn{\Pi',S'}}
    =   \anticommutator{\wh B\fn{\Pi,S}}{\wh B^\dagger\fn{\Pi',S'}}
 &=	\dBK{\Pi,S\md\Pi',S'}\wh1,\\
\tx{others}
	&=	0,
	\label{wave-packet commutator}
}
where $\wh1$ denotes the identity operator in the whole Fock space, and $\dBK{\Pi,S|\Pi',S'}$ is the inner product of the Lorentz covariant wave packets, given in Eq.~\eqref{inner product of Lorentz covariant wave pacekt}.

Finally, the free spinor field can be expanded as
\al{
\wh\psi\fn{x}
	&=	\sum_S\int\df^{2d}\Pi_{\psi}\sqbr{
			U(x,\Pi,S)\wh A\fn{\Pi,S}
			+V(x,\Pi,S)\wh B^\dagger\fn{\Pi,S}
			},
}
where the Dirac spinor wave functions are given by
\al{
U(x,\Pi,S) &= \sum_s\int{\df^d\bs p\ov2p^0} u\fn{p,s}{e^{ip\cdot x}\ov\pn{2\pi}^{d\ov2}} \dBK{p,s\md\Pi,S}\nn
		&= {1\ov2m}{N_\psi\ov N_\phi}\dB{x}(-i\slashed{\h p}+m)\dK{\Pi} u\fn{P,S}\nn
		&= {1\ov2}{N_\psi m^{d-1}\ov\sqrt{2\pi}}
		\pn{-i\slashed{\xi}{K_{d+1\ov2}\fn{\abb\xi}\ov\abb\xi^{d+1\ov2}}
		+{K_{d-1\ov2}\fn{\abb\xi}\ov\abb\xi^{d-1\ov2}}}u(P,S),\\
V(x,\Pi,S) &= \sum_s\int{\df^d\bs p\ov2p^0} v\fn{p,s}{e^{-ip\cdot x}\ov\pn{2\pi}^{d\ov2}} \dBK{\Pi,S\md p,s}\nn
		&=C U^*(x,\Pi,S),
  \label{U and V}
}
in which we have used the scalar wave function~\eqref{scalar wave-packet function}. Here, $\abb\xi$ and $\xi^\mu$ are given in Eq.~\eqref{xi defined} and below it, respectively.

The normalization conditions of these Dirac spinors are
\al{
&\int \frac{d^{d+1} X}{(2\pi)^d} \delta(N\cdot X+T) \ol{U}(x,\Pi,S) ~U(x,\Pi,S')
	= 2m \delta_{SS'}, \nn
&\int \frac{d^{d+1} X}{(2\pi)^d} \delta(N\cdot X+T) \ol{V}(x,\Pi,S) ~V(x,\Pi,S') = -2m \delta_{SS'}
\label{normalization of U and V}
}
where we used Eqs.~\eqref{normalization of u and v} and~\eqref{plane-wave with spin completeness}. The normalization is as same as the case of plane waves \eqref{normalization of u and v}, except for the integration of $X$.

Next, the completeness relations can be computed by
\al{
&\sum_S \int \frac{d^{d+1} X}{(2\pi)^d} \delta(N\cdot X+T) U(x,\Pi,S) ~\ol{U}(x,\Pi,S)
	= -i \slashed{P} \mc M_\psi +m, \nn
&\sum_S \int \frac{d^{d+1} X}{(2\pi)^d} \delta(N\cdot X+T) V(x,\Pi,S) ~\ol{V}(x,\Pi,S)
	= -i \slashed{P} \mc M_\psi -m,
 \label{UUbar and VVbar}
}
where we have used Eqs.~\eqref{completeness of spinor}, \eqref{pslashexpectation} and \eqref{pslashexpectation2}. These relations are similar to that of plane waves~\eqref{completeness of spinor}, except for the integration of $X$ and factor $\mc M_\psi$ on the right-hand side. 

\section{Energy, momentum, and charge}
\label{Energy, momentum, and charge}
In this section, we rewrite well-known operators in QFT, i.e.\ the total Hamiltonian, momentum, and charge operators, in the language of the spinor wave packet. Since the wave packet is not the momentum eigenstate, the total Hamiltonian and momentum operators cannot be diagonalized in the wave packet basis. However, the zero-point energy can be described in a fully Lorentz invariant manner using this basis. In Appendix~\ref{scalar expression}, we also show the corresponding expressions for the scalar wave packet.

First, let us consider the convergent part of the total Hamiltonian and momentum operators. In the momentum space, these operators are given by
\al{
\wh P^{\mu}
	:=	\int{\df^d\bs p\ov2p^0}\sum_s p^{\mu}\pn{\wh\alpha^\dagger\fn{p,s}\wh\alpha\fn{p,s}+\wh\beta^\dagger\fn{p,s}\wh\beta\fn{p,s}}.
}
Putting Eqs.~\eqref{alpha in terms of A for spinor} and ~\eqref{beta in terms of B for spinor} into the above expression, we get
\al{
\wh P^\mu
	&=	\sum_{S,S'} \int\df^{2d}\Pi_{\psi}\int\df^{2d}\Pi'_{\psi}\,\pn{\wh A^\dagger\fn{\Pi}\wh A\fn{\Pi'}+\wh B^\dagger\fn{\Pi}\wh B\fn{\Pi'} }\dBK{\h p^\mu}_{(\Pi,S),(\Pi',S')},
		\label{P hat  spinor given}
}
where
\al{
\dBK{\h p^\mu}_{(\Pi,S),(\Pi',S')}
	&:=	\sum_s \int{\df^d\bs p\ov2p^0}\dBK{\Pi,S\md p,s}p^\mu\dBK{p,s\md \Pi',S'}.
}
We see that the total Hamiltonian and momentum operators are not diagonal on the wave packet basis, unlike the plane-wave eigenbasis.

Let us discuss the divergent part of this operator, coming from the zero-point energy:
\al{
\wh P^{\mu}_{\mathrm{zero}}
	:=	\sum_s \int{\df^d\bs p\ov2p^0}(-p^{\mu})\anticommutator{\wh \beta(p,s)}{\wh \beta^\dagger(p,s)}.
}
Similarly as above, putting Eq.~\eqref{beta in terms of B for spinor} into this commutator, we obtain 
\al{
\wh P^{\mu}_{\mathrm{zero}}
	&=\sum_{S,S'} \int\df^{2d}\Pi_{\psi}\int\df^{2d}\Pi'_{\psi}\,\dBK{-2\h p^{\mu}}_{(\Pi,S),(\Pi',S')} \dBK{\Pi',S'\md\Pi,S}\wh 1\nn
	&=\sum_{S} \int\df^{2d}\Pi_{\psi}\,\dBK{-2\h p^{\mu}}_{\psi}\wh 1\nn
	&= \sum_S \int\frac{\df^{d+1}X}{(2 \pi)^d}{\df^d\bs P\ov2P^0}\pn{2P \cdot N}\,\delta\fn{N\cdot X+T} P^{\mu} \,\wh 1.
	\label{expression for P0 hat mu for spinor}
}
where we have used the completeness relation~\eqref{completeness of spin wave packet} in the second line and, in the last line, the expectation value~\eqref{momentum expectation for spinor} and the Lorentz-invariant phase-space volume element~\eqref{Lorentz invariant measure for spinor}.

Let the time-like normal vector $N^\mu$ and $d$ spacelike vectors $N_{\perp i}$ ($i=1,\dots,d$) compose an orthonormal basis: $N\cdot N_{\perp i}=0$ and $N_{\perp i}\cdot N_{\perp j}=\delta_{ij}$ such that we can decompose $P^{\mu}$ into the components parallel and perpendicular to $N^\mu$,
\al{
P^{\mu}=- \pn{P \cdot N}N^{\mu}+\sum_i \pn{P \cdot N_{\perp i}} N_{\perp i}^{\mu}.
    \label{P decomposed}
}
When we put this into Eq.~\eqref{expression for P0 hat mu for spinor}, the perpendicular components vanish in a regularization scheme that makes Lorentz covariance manifest, namely in the dimensional regularization:
\al{
\int{\df^d\bs P\ov2P^0} \pn{P \cdot N_{\perp i}}  \pn{P \cdot N}
	&= \int{\df^d\bs P\ov2P^0} P^{\mu}P^{\nu} N_{\mu} N_{\perp i\,\nu}\nn
	&= \int{\df^d\bs P\ov2P^0} \frac{1}{d+1}\pn{P\cdot P} \pn{N \cdot N_{\perp i}}\nn
	&=0.
 \label{dropping out orthogonal}
}
Therefore, the divergent part $P^{\mu}_{\mathrm{zero}}$ has only one independent component, which can be interpreted as zero-point energy, defined in a manifestly Lorentz-invariant fashion:
\al{
E_{\mathrm{zero}} 
	&:= -N_\mu P^{\mu}_{\mathrm{zero}} \nn
	& = \sum_S\int\frac{\df^{d+1}X}{(2 \pi)^d}{\df^d\bs P\ov2P^0}\pn{-2}\pn{P \cdot N}^2\,\delta\fn{N\cdot X+T},
		\label{final expression of E0 for spinor}
}
where $P^{\mu}_{\mathrm{zero}}$ is the coefficient in front of $\wh1$ in the right-hand side of Eq.~\eqref{expression for P0 hat mu for spinor}.
We note that the zero-point energy should be a scalar as we have shown, otherwise, an infinite momentum would appear from a Lorentz transformation.

Physically, we would expect that the zero-point energy is independent of the choice of spacelike hyperplane $\Sigma_{N,T}$.
We can show it by exploiting the Lorentz invariance of the expression~\eqref{final expression of E0 for spinor} by choosing $N^{\mu}=\ell^\mu$ ($=\pn{1,\bs 0}$), without loss of generality. Then, the zero-point energy reduces to the well-known form:
\al{
E_{\mathrm{zero}} =\sum_S\int_{X^0=T}{\df^d \bs X\,\df^d\bs P\ov\pn{2\pi}^d} \pn{-P^{0}}.
}
It is remarkable that this zero point energy of the Dirac spinor is exactly $-4$ times that of a real scalar, shown in Eq.~\eqref{zero point energy for scalar} in Appendix, although the expression of momentum expectation values in Eqs.~\eqref{momentum expectation for spinor} and \eqref{momentum expectation} are completely different between the spinor and scalar. 
The factor~$4$ is the number of degrees of freedom, and the negative sign cancels the bosonic contribution in a supersymmetric theory.

Next, we consider the following charge operator
\al{
\wh Q
	:=	\int{\df^d\bs p\ov2p^0}\sum_s\sqbr{-\wh\alpha^\dagger\fn{p,s}\wh\alpha\fn{p,s}+\wh\beta^\dagger\fn{p,s}\wh\beta\fn{p,s}}
}
Substituting Eq.~\eqref{alpha in terms of A for spinor} into the above expression, we obtain
\al{
\label{spinor wave packet charge}
\wh Q 
=	\sum_S\int\df^{2d}\Pi_{\psi} \sqbr{-\wh A^\dagger\fn{\Pi,S}\wh A\fn{\Pi,S}+\wh B^\dagger\fn{\Pi,S}\wh B\fn{\Pi,S}},
}
where we have used
\al{
&\wh A\fn{{\Pi,S}}
	=	\sum_S\int\df^{2d}\Pi'_{\psi}\dBK{\Pi,S\md\Pi',S'}\wh A\fn{{\Pi',S}},\nn
&\wh B\fn{{\Pi,S}}
	=	\sum_S\int\df^{2d}\Pi'_{\psi}\dBK{\Pi',S'\md\Pi,S} \wh B\fn{\Pi',S},
	\label{expansion of X P creation for spinor}
}
which follows from Eq.~\eqref{completeness of spin wave packet}. This expression Eq.~\eqref{spinor wave packet charge} means that the creation operators $A^\dagger\fn{\Pi,S}$ and $B^\dagger\fn{\Pi,S}$ create the wave packet with charge $-1$, and $+1$ respectively. In fact,
\al{
\wh Q \wh A^\dagger\fn{\Pi,S}\ket{0} &=\anticommutator{\wh Q}{\wh A^\dagger\fn{\Pi,S}}\ket{0}\nn
						&=-\sum_{S'}\int\df^{2d}\Pi'_{\psi}\, \wh A^\dagger\fn{\Pi',S'}\dBK{\Pi',S'\md\Pi,S}\ket{0}\nn
						&= -\wh A^\dagger\fn{\Pi,S}\ket{0},
}
is valid. Here we have used Eq.~\eqref{expansion of X P creation for spinor} in the last line.

\section{Summary and discussion}
In this paper, we have proposed fully Lorentz-covariant wave packets with spin. In the conventional definition of the wave packet, spin dependence of the wave function in the momentum space is just given by Kronecker delta, $\delta_{sS}$, and such a wave packet with spin transforms under Lorentz transformation mixing wave-packet states that have different centers of momentum and position. Our proposal overcomes this difficulty.

We have also proven that these wave packets form a complete basis that spans the spinor one-particle subspace in the manifestly Lorentz-invariant fashion.
Generalizing this completeness relation to the whole Fock space, we have shown that the creation and annihilation operators of plane waves can be expanded by that of these wave packets. This relation leads to the expansion of the spinor field in a Lorentz covariant manner. In addition to this, we have expressed the well-known operators in a wave packet basis: the total Hamiltonian, momentum, and charge operators. In particular, we have given the Lorentz covariant expression of zero point energy, in terms of centers of momentum and position of this wave packet.

In the following, we will comment on several future directions. First, as mentioned in the Introduction, the novel Lorentz-covariant basis we propose will be useful in handling wave packet quantum field theory~\cite{Ishikawa:2018koj,Ishikawa:2020hph,Ishikawa:2021bzf,Edery:2021thf}, which has previously relied on the saddle-point approximation for computing position-momentum integrals. For future work, we aim to go beyond the leading-order computation and properly account for the effect of wave packet spreading. To achieve these goals, we require an analytical approach without approximations, which is feasible within our formulation.

Second, neutrino oscillation inherently requires a wave-packet formulation, and our newly defined wave packet could significantly impact this field. Previous analyses have primarily used Gaussian wave packets, without considering the spinorial structure—a key focus of this paper. For instance, Giunti's seminal work~\cite{Giunti:1997wq} addresses neutrino decoherence in quantum mechanics. Akhmedov and Smirnov have also extensively studied neutrino wave packets~\cite{Akhmedov:2009rb,Akhmedov:2022bjs}, exploring phenomenological aspects and scattering processes. Many other groups have contributed to this field as well~\cite{Cardall:1999ze,Akhmedov:2008jn,Kopp:2009fa,Naumov:2009zza,Akhmedov:2010ua,Akhmedov:2010ms,Naumov:2010um,Wu:2010yr,Akhmedov:2012uu,Naumov:2013uia,Blasone:2021cau,Naumov:2020yyv,Naumov:2022kwz,Mitani:2023hpd}. However, these studies do not fully address the spin aspects of neutrinos. Our proposed spinor wave packet framework, which is Lorentz-covariant, may yield novel predictions for neutrino mixing phenomenology and provide a deeper understanding by incorporating the spinorial structure into the wave-packet formulation.

It may also be interesting to consider Bell's inequality in the context of this wave packet state. Another possible application might involve vortex collisions; see, e.g., Ref.~\cite{Ivanov:2022jzh} for a review.

\subsection*{Acknowledgement}
We thank Ryusuke Jinno for a useful comment.
This work is supported in part by the JSPS KAKENHI Grant Nos.~19H01899, 21H01107~(K.O.), and 22KJ1050 (J.W.).

\appendix
\section*{Appendix}
\section{``Slanted'' foliation}\label{``Slanted'' foliation}
In this appendix, we briefly introduce ``slanted'' foliation which is necessary to write down the completeness relation of Lorentz-invariant wave packets in the fully Lorentz-invariant manner.

\begin{figure}[tp]
\hfill
		\includegraphics[height=.23\textheight]{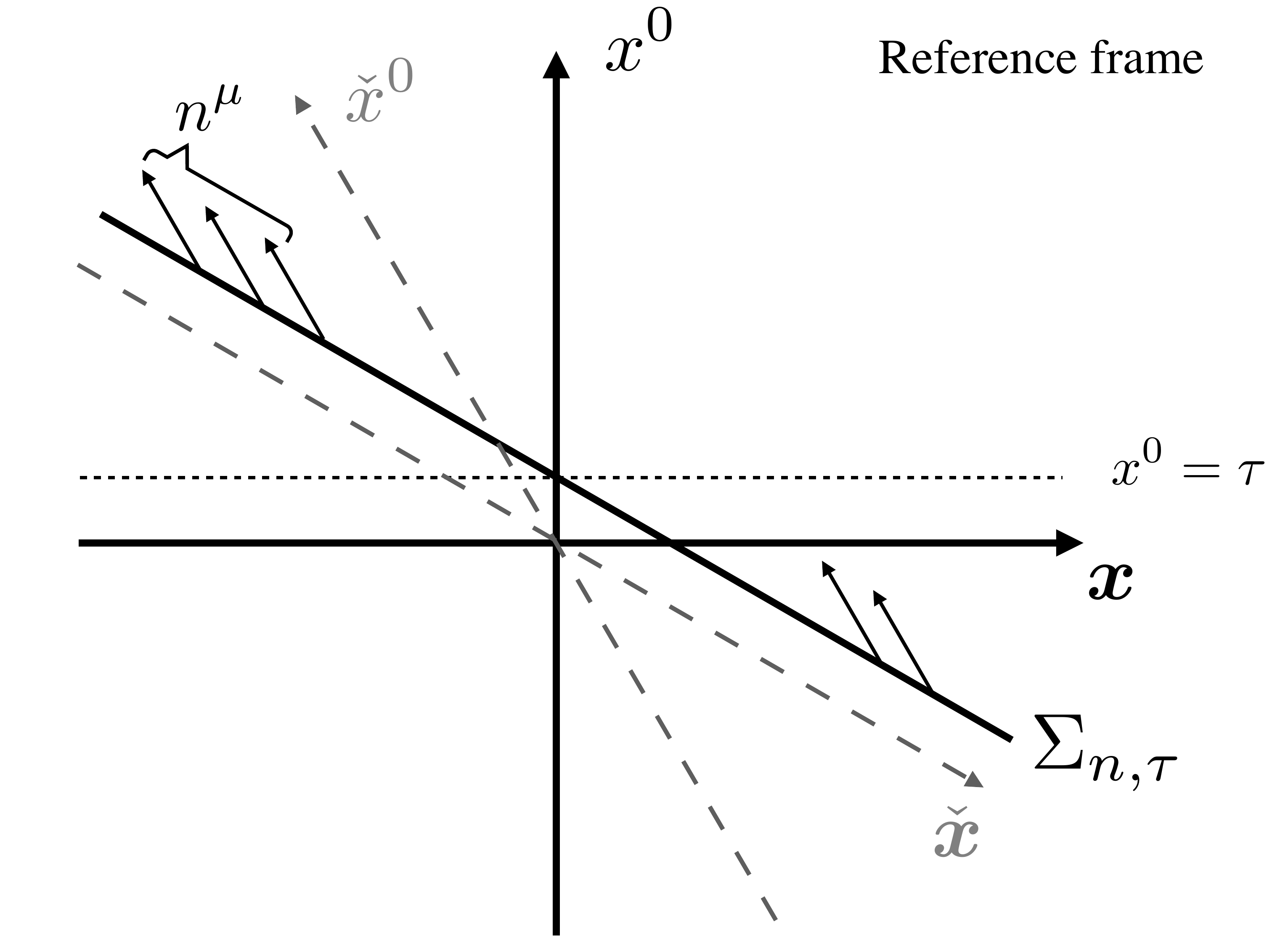}
\hfill
		\includegraphics[height=.23\textheight]{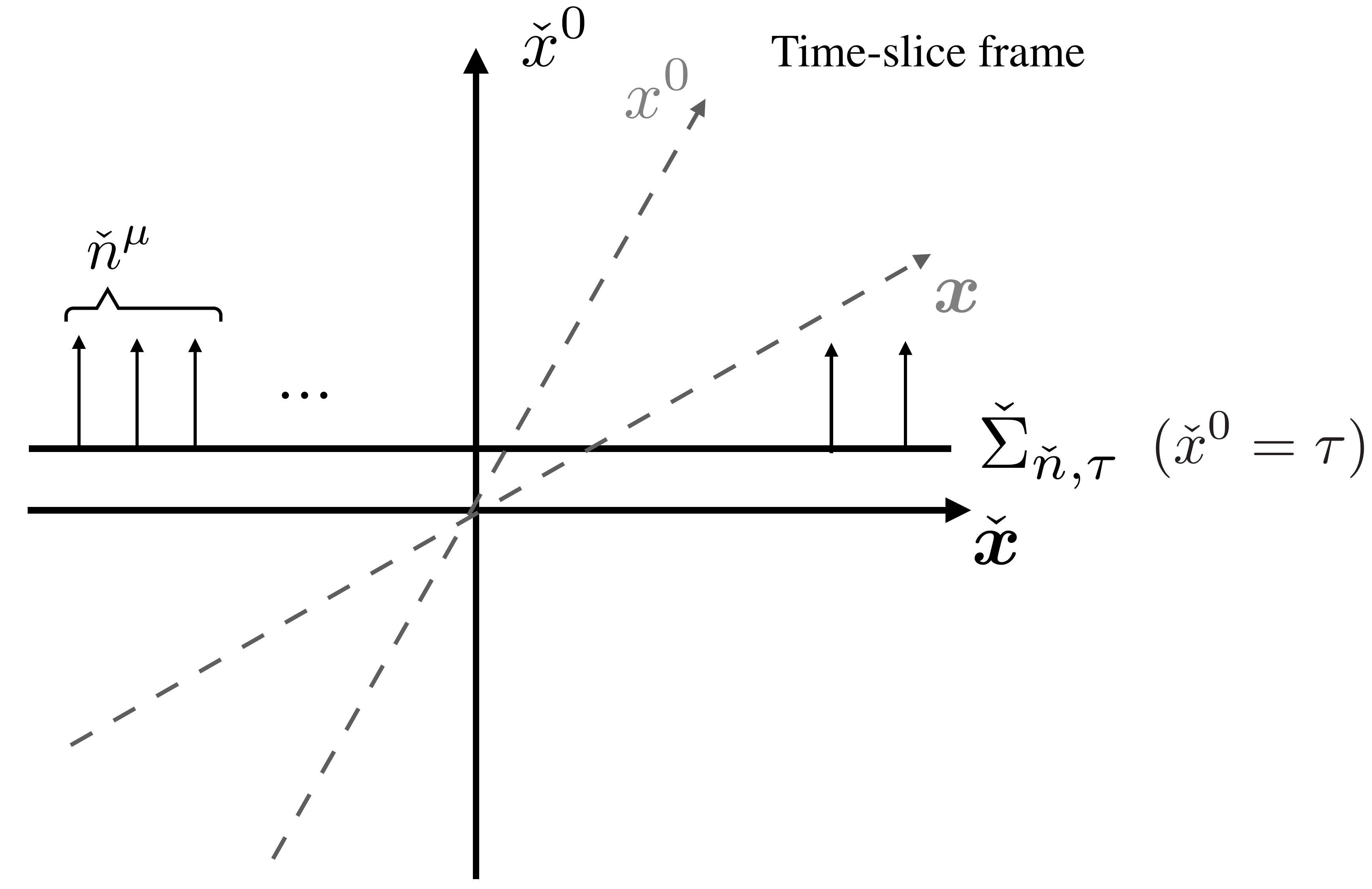}
\hfill\mbox{}\\
	\caption{$\Sigma_{n,\tau}$ in $x$ frame (left) and $\ch\Sigma_{\ch n,\tau}(=\Sigma_{n,\tau})$ in $\check{x}$ frame (right)}
	\label{x and x check figures}
\end{figure}

Let us consider the following spacelike hyperplane:
\al{
\Sigma_{n,\tau}
	&:=	\Set{x\in\mbb R^{1,d}|n\cdot x+\tau=0},
		\label{hyperplane Sigma defined}
}
where $n$ is an arbitrary fixed vector that is timelike-normal $n^2=-1$ and is future-oriented $n^0>0$, namely $n^0=\sqrt{1+\bs n^2}$,
and $\tau\in\mathbb R$ parametrizes the foliation.
Physically, $n$ is the normal vector to the hyperplanes and $\tau$ is the proper time for this foliation.
A schematic figure is given in the left panel in Fig.~\ref{x and x check figures}. We can generalize the equal-time foliation of whole Minkowski space $\mbf M^{d+1}$ to a general foliation by set $\ms F_n=\Set{\Sigma_{n,\tau}}_{\tau\in\mathbb R}$ of these spacelike hyperplanes.

In general, we may parametrize a component of $n$ in the reference frame as the following linear combination:
\al{
n^\mu	&=	\ms L^\mu{}_\nu\fn{\bs n}\sv^\nu,
}
where the ``standard vector'' $\sv$ is defined to be
\al{
\pn{\sv^\mu}_{\mu=0,\dots,d}=\pn{1,\bs 0}
\label{ell given}
}
in any frame\footnote{
In the language of differential geometry, the basis-independent vector is written as $\ch n:=\ch n^\mu\p_\mu$ with $\p_\mu={\p\ov\p x^\mu}$ being the basis vectors in the reference coordinates.
Under the change of basis $\p_\mu\to\p_\mu'=\Lambda_\mu{}^\nu\p_\nu$, where $\p'_\mu:={\p\ov\p x^{\pr\mu}}$, $\ch n$ should remain the same $\ch n\to\ch n^{\pr\mu}\p_\mu'=\ch n^{\pr\mu}\Lambda_\mu{}^\nu\p_\nu\sr!=\ch n^\nu\p_\nu$, that is, $\ch n^{\pr\mu}=\Lambda^\mu{}_\nu\ch n^\nu$.
}
and $\ms L\fn{\bs n}$ is the ``standard boost to the foliation.''
Concretely, for the vector with $n^0=\sqrt{1+\bs n^2}$,
\al{
\ms L\fn{\bs n}
	&=	\bmat{n^0&\bs n^\t\\
			\bs n&\I+\pn{n^0-1}{\bs n\bs n^\t\ov\bs n^2}
			},
			\label{standard boost to the foliation}
}
where $\t$ denotes a transpose, $\I$ is the identity matrix in $d$ dimensions,
$\bs n$ is given in the $n\times1$ matrix representation, and
$\ms L\fn{\bs n}
	=	\bmat{\ms L^\mu{}_\nu\fn{\bs n}}_{\mu,\nu=0,\dots,d}$.
Note that $\ms L^{-1}\fn{\bs n}=\ms L\fn{-\bs n}$.
Now an equal-time hyperplane $x^0=\tau$ in the arbitrary reference frame is written as $\Sigma_{\sv,\tau}$ because $\sv\cdot x+\tau=-x^0+\tau=0$ on it.

For any given foliation $\ms F_n$, we may Lorentz-transform from the reference frame $x$ to the ``time-slice'' frame $\ch x$ that gives $\ch n^\mu=\sv^\mu$:
\al{
x^\mu	&\to	\ch x^\mu=	\ms L_\nu{}^\mu\fn{\bs n}x^\nu,\\
n^\mu	&\to	\ch n^\mu=	\ms L_\nu{}^\mu\fn{\bs n}n^\nu\quad(=\sv^\mu),
}
where we used $\pa{\ms L^{-1}}^\mu{}_\nu=\ms L_\nu{}^\mu$ as usual.
In the time-slice coordinate system $\ch x$, the same plane is written as
\al{
\ch\Sigma_{\ch n,\tau}:=\Set{\ch x=\ms L^{-1}\fn{\bs n}x,\,x\in\mathbb R^{1,d}|\ch n\cdot\ch x+\tau=0}\quad(=\Sigma_{n,\tau}).
	\label{equal-time hyperplane}
}
As said above, since $\ch n\cdot\ch x+\tau=-\ch x^0+\tau=0$ on $\ch\Sigma_{\ch n,\tau}$, they are equal-time hyperplanes parametrized by $\tau\in\mathbb R$ in the $\ch x$ coordinate system.
A schematic figure is given in the right panel in Fig.~\ref{x and x check figures}.

\section{Wigner representation}\label{Wigner section}
In this appendix, we briefly review the Wigner representation in the case of massive one-particle state $\dK{\bs p,s}$ to spell out our notation; see e.g.\ Ref.~\cite{Weinberg:1995mt} for more details. 
Here and hereafter, we neglect the label for the particle and anti-particle since it is irrelevant for the current discussion.

The Poincar\'e transformation on a plane-wave state can be written as
\al{
\dK{p,s}
	&\to	\wh U\fn{\Lambda,a}\dK{p,s}
}
where
\al{
\wh U\fn{\Lambda,a}
	&=	e^{-ia\cdot\wh{\mc P}}\wh U\fn{\Lambda}
	=	e^{ia^0\wh H-i\bs a\cdot\wbh{\mc P}}\wh U\fn{\Lambda},
}
in which $\wh{\mc P}=\pn{\wh{\mc P}^0,\wbh{\mc P}}$ is the generator of the spacetime translation.
Since the translational part is the same as the scalar case, we concentrate on the Lorentz transformation.

Without loss of generality, we can choose $s$ to be the  of the particle in its rest frame:
\al{
\dK{p,s}
	&=	\wh U\Fn{L\fn{p}}\dK{\bs 0,s},
}
consistently with the definition~\eqref{bases inner product} as we will see below.
Here, $s$ is the spin eigenvalue for the rotation in, say, $x^1$-$x^2$ plane in the rest frame and the standard boost $L\fn{p}$ is defined by
\al{
p^\mu
	&=:	L^\mu{}_\nu\fn{p}m\ell^\nu,
}
in which $\ell^\mu$ is given in Eq.~\eqref{ell given}.
Concretely, the standard boost to $p$ can be written in terms of the ``standard boost to a foliation''~\eqref{standard boost to the foliation} as\footnote{
In general, these two are different concepts, $L\fn{\bs p}\neq\ms L\fn{\bs n}$, since $\bs p$ and $\bs n$ are different.
}
\al{
L\fn{p}
	&=	\ms L\fn{p/m}.
}

Since $\dK{p,s}$ has an internal degree of freedom $s$, the Lorentz group representation for this state could be nontrivial. To deal with this, we introduce the well-known procedure, Wigner representation. 
First, under the Lorentz transformation, the plane-wave basis transforms as
\al{
\wh U\fn{\Lambda}\dK{p,s}
	&=	\wh U\fn{\Lambda}\wh U\Fn{L\fn{p}}\dK{\bs 0,s}
	=	\wh U\Fn{L\fn{\Lambda p}}
		\wh U\fn{L^{-1}\fn{\Lambda p}}
		\wh U\fn{\Lambda}\wh U\Fn{L\fn{p}}\dK{\bs 0,s}\nn
	&=	\wh U\Fn{L\fn{\Lambda p}}\wh U\Fn{W\fn{\Lambda,p}}\dK{\bs 0,s},
	\label{Wigner rotation introduced}
}
where
\al{
W\fn{\Lambda,p}
	&:=	L^{-1}\fn{\Lambda p}\Lambda L\fn{p}.
		\label{Winger rotation defined}
}
Here, $W\fn{\Lambda,p}$ is corresponding to the rotation because this transformation does not change the momentum $p$. We call this Wigner rotation in $SO(d)$.\footnote{
We sloppily write $SO(d)$ when it is to be understood as Spin$(d)$.
}
Next, we may always write
\al{
\wh U\Fn{W\fn{\Lambda,p}}\dK{\bs 0,s}
	&=	\sum_{s'}\dK{\bs 0,s'}D_{s's}\Fn{W\fn{\Lambda,p}},
        \label{Wigner rotation}
}
where $D$ is a finite-dimensional unitary representation of $SO(d)$:
\al{
\sum_sD_{s''s}\Fn{W\fn{\Lambda, p}}D^*_{s's}\Fn{W\fn{\Lambda,p}}=\delta_{s's''}.
	\label{D unitarity}
}
Putting Eq.~\eqref{Wigner rotation} into Eq.~\eqref{Wigner rotation introduced}, we obtain Eq.~\eqref{Lorentz transformation on plane-wave momentum state}.

\section{Energy, momentum, and number operator in scalar case}\label{scalar expression}
We give the energy, momentum, and number operators for a real scalar field in terms of the Lorentz-invariant wave-packet basis. First, we briefly review the scalar wave packet in QFT, and then we will show newly-found expressions of these operators in terms of the Lorentz-invariant scalar wave packets.

The free field is usually expressed in the plane wave basis:
\al{
\wh\phi\fn{x}
	&=	\int{\df^d\bs p\ov2p^0}\pn{\wh\alpha\fn{p}{e^{ip\cdot x}\ov\pn{2\pi}^{d/2}}+\wh \alpha^\dagger\fn{p}{e^{-ip\cdot x}\ov\pn{2\pi}^{d/2}}},
		\label{common expression of scalar expansion}
}
where $\wh \alpha\fn{p}$ and $\wh \alpha^\dagger\fn{p}$ are the creation and annihilation operators of the plane waves, which satisfy $\commutator{\wh\alpha\fn{p}}{\wh\alpha^\dagger\fn{p'}}=2p^0\,\delta^d\fn{\bs p-\bs p'}$, etc.

Now, we define a wave-packet creation operator by~\cite{Oda:2021tiv}
\al{
\wh A^{\dagger}\fn{\Pi}\ket{0}:=\dK{\Pi},
}
and an annihilation operator $\wh A\fn{\Pi}$ by its Hermitian conjugate. 
The completeness of the scalar wave packet~\eqref{completeness relation for scalar} leads to the following expansion of the creation and annihilation operators of the plane waves:
\al{
\wh\alpha\fn{p}
	&=	\int\df^{2d}\Pi\dBK{p\md\Pi}\wh A\fn{\Pi},&
\wh\alpha^\dagger\fn{p}
	&=	\int\df^{2d}\Pi\,\wh A^\dagger\fn{\Pi}\dBK{\Pi\md p}.
		\label{alpha in terms of A}
}
Thus, the free scalar field can be expanded as~\cite{Oda:2021tiv}
\al{
\wh\phi\fn{x}
	&=	\int\df^{2d}\Pi\sqbr{
			\dBK{x\md\Pi}\wh A\fn{\Pi}
			+\wh A^\dagger\fn{\Pi}\dBK{\Pi\md x}
			},
}
where the wave function is given in Eq.~\eqref{scalar wave-packet function}.

Now let us rewrite well-known operators in QFT, i.e.\ the total Hamiltonian, momentum, and number operators, into the language of the scalar wave packet.

First, in momentum space, the convergent part of the number operator is described by
\al{
\wh N
	:=	\int{\df^d\bs p\ov2p^0}\wh\alpha^\dagger\fn{p}\wh\alpha\fn{p}.
}
Substituting Eq.~\eqref{alpha in terms of A} into the above expression, we obtain
\al{
\wh N 
&=	\int\df^{2d}\Pi\,\int\df^{2d}\Pi'\,\wh A^\dagger\fn{\Pi}\dBK{\Pi\md\Pi'}\wh A\fn{\Pi'}\nn
&=	\int\df^{2d}\Pi \, \wh A^\dagger\fn{\Pi}\wh A\fn{\Pi}.
    \label{number operator}
}
On the second line, we have used
\al{
&\wh A\fn{\Pi}
	=	\int\df^{2d}\Pi'\dBK{\Pi\md\Pi'}\wh A\fn{\Pi'},
&\wh A^\dagger\fn{\Pi}
	=	\int\df^{2d}\Pi' \wh A^\dagger\fn{\Pi'}\dBK{\Pi'\md\Pi},
	\label{expansion of X P creation}
}
which follows from Eq.~\eqref{completeness relation for scalar}. From Eq.~\eqref{number operator}, we can read off a Lorentz-covariant number-density operator in the $2d$-dimensional phase space:
\al{
\wh{\mc{N}} = \wh A^\dagger\fn{\Pi}\wh A\fn{\Pi}.
}

We now consider the divergent part of the plane-wave number operator, coming from the zero-point oscillation:
\al{
\wh N_{\mathrm{zero}}
	:=	\int{\df^d\bs p\ov2p^0} \frac{1}{2}\commutator{\wh \alpha\fn{p}}{\wh \alpha^\dagger\fn{p}}.
}
Putting Eq.~\eqref{alpha in terms of A} into the above expression, we obtain
\al{
\wh N_{\mathrm{zero}}
	&=	\int\df^{2d}\Pi\,\int\df^{2d}\Pi'\, \frac{1}{2}\dBK{\Pi\md\Pi'}\dBK{\Pi'\md\Pi}\wh 1\nn
	&=	\frac{1}{2} \int\df^{2d}\Pi \, \wh 1.
}
Therefore, including the divergent part, the number-density operator can be described by
\al{
\wh{\mc{N}} + \wh{\mc{N}}_{\mathrm{zero}} := \wh A^\dagger\fn{\Pi}\wh A\fn{\Pi} + \frac{1}{2}
}
From this expression, it can be interpreted that there is one zero-point oscillation per $2d$-dimensional phase space volume.

Next, we consider the convergent part of the total Hamiltonian and momentum operators. In the momentum space, these operators are given by
\al{
\wh P^{\mu}
	:=	\int{\df^d\bs p\ov2p^0}p^{\mu}\wh\alpha^\dagger\fn{p}\wh\alpha\fn{p}.
}
Putting Eq.~\eqref{alpha in terms of A} into the above expression, we get
\al{
\wh P^\mu
	&=	\int\df^{2d}\Pi\int\df^{2d}\Pi'\,\wh A^\dagger\fn{\Pi}\wh A\fn{\Pi'}\dBK{\h p^\mu}_{\Pi,\Pi'},
		\label{P hat given}
}
where $\dBK{\h p^\mu}_{\Pi,\Pi'}$ is given in Eq.\eqref{matrix element of p}.
We see that the total Hamiltonian and momentum operators are not diagonal on the wave packet basis, unlike the plane-wave eigenbasis.

Now, let us discuss the divergent part of this operator, coming from the zero-point energy:
\al{
\wh P^{\mu}_{\mathrm{zero}}
	:=	\int{\df^d\bs p\ov2p^0}p^{\mu}\frac{1}{2}\commutator{\wh \alpha\fn{p}}{\wh \alpha^\dagger\fn{p}}.
}
Putting Eq.~\eqref{alpha in terms of A} into the above commutator, we obtain 
\al{
\wh P^{\mu}_{\mathrm{zero}}
	&=\frac{1}{2}\int\df^{2d}\Pi\,\fn{\Pi}\int\df^{2d}\Pi'\,\dBK{\h p^{\mu}}_{\Pi,\Pi'} \dBK{\Pi'\md\Pi}\wh 1\nn
	&=\frac{1}{2}\int\df^{2d}\Pi\,\dBK{\h p^{\mu}}_{\phi}\wh 1,\nn
	&= \int\frac{\df^{d+1}X}{(2 \pi)^d}{\df^d\bs P\ov2P^0}\pn{-P \cdot N}\,\delta\fn{N\cdot X+T} P^{\mu} \,\wh 1\nn
	&= \int\frac{\df^{d+1}X}{(2 \pi)^d}{\df^d\bs P\ov2P^0}\pn{P \cdot N}^2\,\delta\fn{N\cdot X+T} N^{\mu} \,\wh 1.
	\label{expression for P0 hat mu}
}
where we have used the completeness relation~\eqref{completeness relation for scalar} in the second line,
the formula~\eqref{momentum expectation} in the third line,
and
the same argument as in Eq.~\eqref{dropping out orthogonal} in the last line.
It is noteworthy that the result becomes the same as in the spinor case~\eqref{expression for P0 hat mu for spinor} up to the factor $-4$.

 We may define the zero-point energy in a manifestly Lorentz-invariant fashion:
\al{
E_{\mathrm{zero}} 
	&:= -N_\mu P_{\mathrm{zero}}^\mu \nn
	& = \int\frac{\df^{d+1}X}{(2 \pi)^d}{\df^d\bs P\ov2P^0}\pn{P \cdot N}^2\,\delta\fn{N\cdot X+T},
		\label{final expression of E0}
}
where $P_\mathrm{zero}^\mu$ is the coefficient of $\wh1$ in the right-hand side of Eq.~\eqref{expression for P0 hat mu}.

Physically, we expect that the zero-point energy should be independent of the choice of the spacelike hyperplane $\Sigma_{N,T}$.
We can show it by exploiting the Lorentz invariance of the expression~\eqref{final expression of E0} to choose $N^{\mu}=\ell^\mu$ ($=\pn{1,\bs 0}$), without loss of generality. Then, the zero-point energy reduces to the well-known form:
\al{
\label{zero point energy for scalar}
E_{\mathrm{zero}}=\int_{X^0=T}{\df^d \bs X\,\df^d\bs P\ov\pn{2\pi}^d} \frac{1}{2}P^{0}.
}

\section{Conversion of metric convention}\label{metric conversion}
We list the corresponding equations in the almost-minus metric signature $\pn{+,-,\dots,-}$ to those in the main text.
In this paragraph only, we tentatively put the subscript ``Main'' and ``AppD'' on those in the main text and Appdendix~D, respectively.
The metric sign is reverted: $\eta_\tx{AppD}=\diag\fn{1,-1,\dots,-1}$, and $\beta=\gamma^0_\tx{AppD}$; see footnote~\ref{metric convention footnote}.
The gamma matrices are related by $\gamma^\mu_\tx{Main}=-i\gamma^\mu_\tx{AppD}$. The Clifford algebra remains the same: $\{\gamma_\tx{AppD}^\mu,\gamma_\tx{AppD}^\nu\}=2\eta^{\mu\nu}_\tx{AppD}I$.

\subsection*{Section~\ref{introduction}}
Eq.~\eqref{main proposal} becomes
\al{
\dBK{p,s\md X,P,S;\sigma}
    &\propto
        e^{ip\cdot\pn{X+i\sigma P}}\ol u\fn{p,s}u\fn{P,S}\nn
    &\quad=   -e^{ip\cdot\pn{X+i\sigma P}}\ol v\fn{P,S}v\fn{p,s}.
}

\subsection*{Section~\ref{Lorentz-covariant spinor wave packet}}
Eq.~\eqref{Dirac field expansion plane wave} becomes
\al{
\wh\psi\fn{x}
	&=	\sum_s \int{\df^d\bs p\ov2p^0}\pn{
			u(p,s){e^{-ip\cdot x}\ov\pn{2\pi}^{d/2}}\wh\alpha\fn{p,s}
			+v(p,s){e^{ip\cdot x}\ov\pn{2\pi}^{d/2}}\wh \beta^\dagger\fn{p,s}}.
}
Eq.~\eqref{Dirac equation} becomes
\al{
&(\slashed{p}-m)u(p,s)=0,\nn
&(\slashed{p}+m)v(p,s)=0.
}
Eq.~\eqref{completeness of spinor} becomes
\al{
\sum_s u(p,s)\ol{u}(p,s)
	&=	\slashed{p}+m,\nn
\sum_s v(p,s)\ol{v}(p,s)
	&=	\slashed{p}-m.
}
Eq.~\eqref{normalization of u and v} is unchanged; hereafter, we do not mention unchanged equations.
Eq.~\eqref{LIWPdefinition} becomes
\al{
\dBK{p\md\Pi}
	&:=	N_\phi e^{ip\cdot\pn{X+i\sigma P}}.
}

The ``Lorentzian norm'' becomes
\al{
\abb V:=\sqrt{V^2},
}
and with the same definition $\xi^\mu:=m\sqbr{\sigma P^\mu+i\pn{x-X}^\mu}$ and $\Xi^\mu:=m\sqbr{\sigma\pn{P+P'}^\mu+i\pn{X-X'}^\mu}$,
Eqs.~\eqref{xi defined} and \eqref{Xi parameter} becomes
\al{
\abb\xi
	&=	m\sqrt{\sigma^2m^2-\pn{x-X}^2+2i\sigma P\cdot\pn{x-X}},
		\\
\abb\Xi
	&=	m\sqrt{-\Pn{\pn{X-X'}-i\sigma\pn{P+P'}}^2}.
}
Eq.~\eqref{momentum covariance} becomes
\al{
\dBK{\h p^\mu\h p^\nu}_{\phi}
	&=	{K_{d+3\ov2}\fn{2\sigma m^2}\ov K_{d-1\ov2}\fn{2\sigma m^2}}
                P^\mu P^\nu
        -{\mc M_\phi\ov2\sigma}\eta^{\mu\nu}.
}
The spacelike hyperplane becomes $\Sigma_{N,T}=\Set{X|N\cdot X-T=0}$.
Eq.~\eqref{Lorentz invariant measure} becomes
\al{
\int\df^{2d}\Pi_{\phi}
	&:=	{1\ov\mc M_\phi}
        \int{\df^d\Sigma^\mu_X\ov\pn{2\pi}^d}\pn{2P_\mu}{\df^d\bs P\ov2P^0}.
}
Eq.~\eqref{volume element vector} becomes
\al{
\df^d\Sigma^\mu_X
	&:=	\df^{d+1}X\,\delta\fn{N\cdot X-T}N^\mu.
}

Eq.~\eqref{general choice} reads
\al{
\dBK{p,s,N\md\Pi,S,N'}
    :=   N_\psi e^{ip\cdot\pn{X+i\sigma P}}M_{sS}\fn{p,P,N} \delta_{NN'}.
}
The charge conjugation matrix becomes $C=-i\gamma^2$ with $C^*=C^\t=C^\dagger=C$.
Eq.~\eqref{normalization for particle} becomes
\al{
\dBK{\Pi,S\md\Pi,S}
	&=	{N_\psi^2\ov\pn{2m}^2}\ol u\fn{P,S}\pn{\int{\df^d\bs p\ov2p^0}{\dBK{\Pi\md p}\dBK{p\md\Pi}\ov N_\phi^2}
		\pn{\slashed p+m}}
		u\fn{P,S}\nn
	&=	{N_\psi^2\ov\pn{2m}^2}\ol u\fn{P,S}{\dBK{\slashed{\h p}+m}_\phi\ov N_\phi^2}
		u\fn{P,S}\nn
    &\pn{
        ={N_\psi^2\ov\pn{2m}^2}\ol v\fn{P,S}{\dBK{\slashed{\h p}-m}_\phi\ov N_\phi^2}
		v\fn{P,S}
    }.
}
Eq.~\eqref{pslashexpectation} becomes
\al{
\dBK{\slashed{\h p}\pm m}_{\phi}
    =   \mc M_\phi\slashed P
        \pm m.
}
Eq.~\eqref{inner product of Lorentz covariant wave pacekt} becomes
\al{
\dBK{\Pi,S\md\Pi',S'}
	&=	{N_\psi^2\ov\pn{2m}^2}\ol u\fn{P,S}{\dBK{\slashed{\h p}+m}_{\Pi,\Pi'}\ov N_\phi^2}
		u\fn{P',S'}\delta_{nn'}\nn
	&\pn{=	{N_\psi^2\ov\pn{2m}^2}\ol v\fn{P,S}{\dBK{\slashed{\h p}-m}_{\Pi,\Pi'}\ov N_\phi^2}
		v\fn{P',S'}\delta_{n^\tx{c}n^{\pr\tx{c}}}
  }.
}

Eq.~\eqref{p expectation value by psi} becomes
\al{
\dBK{\h p^{\mu}}_{\psi} 
	&= {N_\psi\ov 4m^2}\ol{u}(P,S){\dBK{\h p^{\mu}\pn{\slashed{\h p}+m}}_{\phi}\ov N_\phi^2}u(P,S)\ 
 \pn{
    ={N_\psi\ov 4m^2}\ol{v}(P,S){\dBK{\h p^{\mu}\pn{\slashed{\h p}-m}}_{\phi}\ov N_\phi^2}v(P,S)
        }.
}
Eq.~\eqref{pslashexpectation2} becomes
\al{
\dBK{\h p^{\mu}\pn{\slashed{\h p}\pm m}}_{\phi}
    &=  \paren{
			{K_{d+3\ov2}\fn{2\sigma m^2}\ov K_{d-1\ov2}\fn{2\sigma m^2}}
                P^{\mu} \slashed{P}
            +{\mc M_\phi\ov2\sigma}\gamma^{\mu}
			} \pm\mc M_\phi m P^{\mu}.
}
Eq.~\eqref{useful equation} becomes
\al{
{1\ov4m^2}\ol{u}(P,S)\dBK{\h p^{\mu}\pn{\slashed{\h p}+m}}_{\phi}u(P,S)
    &=  {1\ov2}\paren{
			\mc M_\phi +
			{\mc M_\phi\ov2\sigma m^2}
			+
				{K_{d+3\ov2}\fn{2\sigma m^2}\ov K_{d-1\ov2}\fn{2\sigma m^2}}}P^{\mu},\nn
{1\ov4m^2}\ol{v}(P,S)\dBK{\h p^{\mu}\pn{\slashed{\h p}-m}}_{\phi}v(P,S)
    &=  {1\ov2}\paren{
			\mc M_\phi +
			{\mc M_\phi\ov2\sigma m^2}
			+
				{K_{d+3\ov2}\fn{2\sigma m^2}\ov K_{d-1\ov2}\fn{2\sigma m^2}}}P^{\mu}.
}

Eq.~\eqref{Lorentz invariant measure for spinor} becomes 
\al{
\int\df^{2d}\Pi_{\psi}
	&:=
        {1\ov\mc M_\psi}
        \int_{\Sigma_{N,T}}{\df^d\Sigma^\mu_X\ov\pn{2\pi}^d}\pn{2P_\mu}{\df^d\bs P\ov2P^0}\nn	
	&=  {\mc M_\phi\ov\mc M_\psi}
        \int\df^{2d}\Pi_{\phi},
}
Eq.~\eqref{expression of completeness of spin wave packet} becomes
\al{
&\quad{N_\psi^2\ov\mc M_\psi}\int_{\Sigma_{N,T}}{\df^{d+1}X\ov\pn{2\pi}^d}\delta\fn{N\cdot X-T}\pn{2P\cdot N}{\df^d\bs P\ov2P^0}{\dBK{p\md\Pi}\dBK{\Pi\md q}\ov N_\phi^2}&\nn
&\qquad\times \ds{1\ov4m^2}\sum_S \ol u\fn{p,s}u\fn{P,S} \ol u\fn{P,S}u\fn{q,s'}\nn
	&=	2 p^0 \delta^d (\bs{p}-\bs{q})\delta_{ss'}.
}
Eq.~\eqref{lhs eq} becomes
\al{
(\tx{l.h.s.})
    &= {N_{\psi}^2\ov\mc M_\psi^2}\delta^d (\bs{p}-\bs{q})\int{\df^d\bs P\ov2P^0}2P^0{e^{2\sigma P \cdot p}\ov N_\phi^2}
        {\ol u\fn{p,s} \pn{\slashed P+m} u\fn{p,s'}\ov4m^2} \nn
	&= {N_{\psi}^2\ov\mc M_\psi^2}  \delta^d (\bs{p}-\bs{q}) \ol u\fn{p,s}{\dBK{2\h p^0\pn{\slashed{\h p}+m}}_{\phi}\ov N_\phi^2}
		u\fn{q,s'}\nn
	&=2 p^0 \delta^d (\bs{p}-\bs{q}).
} 

\subsection*{Section~\ref{Spinor field expanded by wave packets}}
Eq.~\eqref{U and V} becomes
\al{
U(x,\Pi,S) &= \sum_s\int{\df^d\bs p\ov2p^0} u\fn{p,s}{e^{-ip\cdot x}\ov\pn{2\pi}^{d\ov2}} \dBK{p,s\md\Pi,S}\nn
		&= {1\ov2m}{N_\psi\ov N_\phi}\dB{x}\pn{\slashed{\h p}+m}\dK{\Pi} u\fn{P,S}\nn
		&= {1\ov2}{N_\psi m^{d-1}\ov\sqrt{2\pi}}
		\pn{\slashed{\xi}{K_{d+1\ov2}\fn{\abb\xi}\ov\abb\xi^{d+1\ov2}}
		+{K_{d-1\ov2}\fn{\abb\xi}\ov\abb\xi^{d-1\ov2}}}u(P,S),\\
V(x,\Pi,S) &= \sum_s\int{\df^d\bs p\ov2p^0} v\fn{p,s}{e^{ip\cdot x}\ov\pn{2\pi}^{d\ov2}} \dBK{\Pi,S\md p,s}\nn
		&=C U^*(x,\Pi,S).
}
Eq.~\eqref{normalization of U and V} becomes
\al{
&\int \frac{d^{d+1} X}{(2\pi)^d} \delta(N\cdot X-T) \ol{U}(x,\Pi,S) ~U(x,\Pi,S')
	= 2m \delta_{SS'}, \nn
&\int \frac{d^{d+1} X}{(2\pi)^d} \delta(N\cdot X-T) \ol{V}(x,\Pi,S) ~V(x,\Pi,S') = -2m \delta_{SS'}.
}
Eq.~\eqref{UUbar and VVbar} becomes
\al{
&\sum_S \int \frac{d^{d+1} X}{(2\pi)^d} \delta(N\cdot X-T) U(x,\Pi,S) ~\ol{U}(x,\Pi,S)
	= \slashed{P} \mc M_\psi +m, \nn
&\sum_S \int \frac{d^{d+1} X}{(2\pi)^d} \delta(N\cdot X-T) V(x,\Pi,S) ~\ol{V}(x,\Pi,S)
	= \slashed{P} \mc M_\psi -m.
}

\subsection*{Section~\ref{Energy, momentum, and charge}}
Eq.~\eqref{expression for P0 hat mu for spinor} becomes
\al{
\wh P^{\mu}_{\mathrm{zero}}
	&=\sum_{S,S'} \int\df^{2d}\Pi_{\psi}\int\df^{2d}\Pi'_{\psi}\,\dBK{-2\h p^{\mu}}_{(\Pi,S),(\Pi',S')} \dBK{\Pi',S'\md\Pi,S}\wh 1\nn
	&=\sum_{S} \int\df^{2d}\Pi_{\psi}\,\dBK{-2\h p^{\mu}}_{\psi}\wh 1\nn
	&= \sum_S \int\frac{\df^{d+1}X}{(2 \pi)^d}{\df^d\bs P\ov2P^0}\pn{-2P \cdot N}\,\delta\fn{N\cdot X-T} P^{\mu} \,\wh 1.
}
Eq.~\eqref{P decomposed} becomes
\al{
P^{\mu}=\pn{P \cdot N}N^{\mu}-\sum_i \pn{P \cdot N_{\perp i}} N_{\perp i}^{\mu}.
}
Eq.~\eqref{final expression of E0 for spinor} becomes
\al{
E_{\mathrm{zero}} 
	&:= N_\mu P^{\mu}_{\mathrm{zero}} \nn
	& = \sum_S\int\frac{\df^{d+1}X}{(2 \pi)^d}{\df^d\bs P\ov2P^0}\pn{-2}\pn{P \cdot N}^2\,\delta\fn{N\cdot X-T}.
}

\bibliographystyle{JHEP}
\bibliography{refs}

\end{document}